\documentclass[12pt]{article}
\usepackage[left=2.5cm,top=2.50cm,right=2.5cm,bottom=2.50cm]{geometry}
\usepackage{mathrsfs}
\usepackage{amsmath,amssymb,latexsym,color,cancel,graphicx,bbm,colortbl}
\usepackage[english]{babel}
\usepackage[latin1]{inputenc}
\usepackage{ragged2e}
\begin{document}
\date{}

\title{Coherent states for the two-dimensional Dirac-Moshinsky oscillator coupled to an external magnetic field}
\author{D. Ojeda-Guill\'en$^{a}$,\footnote{{\it E-mail address:} dojedag@ipn.mx}\\ R. D. Mota$^{b}$ and V. D. Granados$^{a}$} \maketitle

\begin{minipage}{0.9\textwidth}
\small $^{a}$ Escuela Superior de F{\'i}sica y Matem\'aticas,
Instituto Polit\'ecnico Nacional, Ed. 9, Unidad Profesional Adolfo L\'opez Mateos, C.P. 07738, M\'exico D. F., Mexico.\\

\small $^{b}$ Escuela Superior de Ingenier{\'i}a Mec\'anica y El\'ectrica, Unidad Culhuac\'an,
Instituto Polit\'ecnico Nacional, Av. Santa Ana No. 1000, Col. San Francisco Culhuac\'an, Delegaci\'on Coyoac\'an, C.P. 04430,  M\'exico D. F., Mexico.\\

\end{minipage}

\begin{abstract}
We show that the $(2+1)$-dimensional Dirac-Moshinsky oscillator coupled to an external magnetic field can be treated
algebraically with the $SU(1,1)$ group theory and its group basis. We use the $su(1,1)$ irreducible representation theory to find
the energy spectrum and the eigenfunctions. Also, with the $su(1,1)$ group basis we construct the
relativistic coherent states in a closed form for this problem.

\end{abstract}

PACS: 02.20.Sv, 03.65.Fd, 42.50.Ar, 03.65.Pm\\
Keywords: Lie algebras, coherent states, Dirac equation, Dirac-Moshinsky oscillator

\section{Introduction}

The study of the non-relativistic quantum harmonic oscillator leads to Schrödinger to discover the coherent states \cite{Scrho}.
The works of Glauber \cite{Glau}, Klauder \cite{Klau} and Sudarshan \cite{Sudar} related the coherent states to the field of quantum optics. Barut and Girardello \cite{BandG} and Perelomov \cite{Perel} extended the idea of harmonic oscillator coherent states (related to the Heisenberg-Weyl group) for
other Lie groups, in particular for the $SU(1,1)$ group. Since its introduction, the coherent states have been obtained for many problems
in physics, as it is shown in references \cite{Klauderlibro,Gazeaulibro}.

On the other hand, the Dirac oscillator was first introduced by Ito \emph{et al.} \cite{Ito} and Cook \cite{Cook} by adding the linear term
$-imc\omega\beta{\mathbf{\alpha}}\cdot \mathbf{r}$ to the relativistic momentum $\textbf{p}$ of the free-particle Dirac equation.
Moshinsky and Szczepaniak reintroduced this problem and called it ``Dirac oscillator". They constructed its explicit solutions \cite{Mos},
showed that its symmetry Lie algebra is $so(4)\oplus so(3,1)$ and found its generators \cite{Mos2}. For the uncoupled equations
the $su(1,1)$ Lie algebra was introduced in reference \cite{Salas}. This problem has the property that,
in the non-relativistic limit, it reduces to the harmonic oscillator plus a spin-orbit coupling term. Since its introduction, the
Dirac-Moshinsky oscillator has been extensively studied and applied in many branches of physics, as can be seen in references
\cite{Delgado}-\cite{deLima}.

The importance of the study of the $(3+1)$-dimensional Dirac-Moshinsky oscillator relies in its application to quark
confinement models in quantum chromodynamics. Moreover, the $(2+1)$-dimensional Dirac-Moshinsky oscillator has been related to quantum optics \cite{Bermudez,Sadurni} via the Jaynes-Cummings and anti-Jaynes-Cummings model \cite{Jaynes}.
Recently, the Dirac-Moshinsky oscillator coupled to an external magnetic field has been studied.
In reference \cite{Mandal} it has been shown that this problem can be mapped to the anti-Jaynes-Cummings model
for arbitrary strength of the magnetic field.

The aim of the present work is to study algebraically the $(2+1)$-dimensional Dirac-Moshinsky oscillator coupled to an
external magnetic field by introducing properly the $SU(1,1)$ group theory and its irreducible unitary representation.
We obtain the energy spectrum, the eigenstates and the radial coherent states in a closed form for this problem.

This work is organized as it follows. In Section $2$, we introduce the Dirac-Moshinsky oscillator
in $2+1$ dimensions coupled to an external magnetic field in polar coordinates. We decouple
the radial equations for the upper and lower wave functions. We notice that each of these equations
can be treated by an $su(1,1)$ Lie algebra realization and obtain the energy spectrum and the eigenfunctions.
In Section $3$, we construct the $SU(1,1)$ Perelomov coherent states for each radial components, and we calculate
the normalized complete radial coherent state. Finally, we give some concluding remarks.

\section{Dirac-Moshinsky oscillator coupled to an external magnetic field.}

The time-independent Dirac equation for the Dirac-Moshinsky oscillator is given by the Hamiltonian \cite{Mos}
\begin{equation}
H_D\Psi=\left[c \mathbf{\alpha} \cdot\left(\mathbf{p}-im\omega \mathbf{r}\beta\right)+mc^2\beta\right]\Psi=E\Psi.
\end{equation}
In the presence of an external uniform magnetic field $\mathbf{B}$, the Dirac Hamiltonian is modified as
\begin{equation}
H_D=c \mathbf{\alpha} \cdot\left[\left(\mathbf{p}-\frac{e\mathbf{A}}{c}\right)-im\omega \mathbf{r}\beta\right]+mc^2\beta,\label{hamc}
\end{equation}
where $\omega$ is the frequency, $\mathbf{r}$ is the position vector of the oscillator and $\mathbf{A}$ is the vector potential
in the symmetric gauge:
\begin{equation}
\mathbf{A}=\left(-\frac{By}{2},\frac{Bx}{2}\right).
\end{equation}
In $2+1$ dimensions Dirac matrices are set in the standard way in terms of Pauli spin matrices
$\alpha_1=\sigma_1, \alpha_2=\sigma_2, \beta=\sigma_3$. In polar coordinates $(r,\varphi)$
the Hamiltonian of equation (\ref{hamc}) takes the form
\begin{equation}
H_D=\begin{pmatrix}
mc^2 & c\hbar e^{-i\varphi}\left(\frac{1}{i}\frac{\partial}{\partial r}-\frac{1}{r}\frac{\partial}{\partial\varphi}\right)+imc\bar{\omega} re^{-i\varphi} \\
c\hbar e^{i\varphi}\left(\frac{1}{i}\frac{\partial}{\partial r}+\frac{1}{r}\frac{\partial}{\partial\varphi}\right)-imc\bar{\omega} re^{i\varphi} & -mc^2 \end{pmatrix},
\end{equation}
where $\bar{\omega}=\omega-\frac{\omega_c}{2}$ and $\omega_c=\frac{|e|B}{mc}$ is the cyclotron frequency. Since the Hamiltonian commutes with the total angular momentum $J_z=L_z+\frac{1}{2}\hbar\sigma_3$, both $H_D$ and $J_z$ have simultaneous eigenfunctions. Thus, with the following ansatz for the
wave function
\begin{equation}
\psi(\mathbf{r})=
\begin{pmatrix}
\psi_1(\mathbf{r}) \\
\psi_2(\mathbf{r}) \end{pmatrix}
=\begin{pmatrix}
iG(r)\frac{1}{\sqrt{2\pi}}e^{il\varphi} \\
-F(r)\frac{1}{\sqrt{2\pi}}e^{i(l+1)\varphi} \end{pmatrix},
\end{equation}
the coupled equations for the radial components $G(r)$ and $F(r)$ are
\begin{equation}
\left[-\frac{d}{dr}-\frac{(l+1)}{r}+\frac{m\bar{\omega}r}{\hbar}\right]F(r)=-\left(\frac{E-mc^2}{c\hbar}\right)G(r),\label{cou1}
\end{equation}
\begin{equation}
\left[-\frac{d}{dr}+\frac{l}{r}-\frac{m\bar{\omega}r}{\hbar}\right]G(r)=\left(\frac{E+mc^2}{c\hbar}\right)F(r).\label{cou2}
\end{equation}
From these expressions we obtain the uncoupled equation for the radial function $G(r)$
\begin{equation}
\left[-\frac{d^2}{dr^2}-\frac{1}{r}\frac{d}{dr}+\frac{l^2}{r^2}-\frac{2m\bar{\omega}(l+1)}{\hbar}+\frac{m^2\bar{\omega}^2r^2}{\hbar^2}
-\left(\frac{E}{\hbar c}\right)^2+\left(\frac{mc}{\hbar}\right)^2\right]G(r)=0.\label{dgr}
\end{equation}
By making the substitution $r\rightarrow bx$, with $b=\sqrt{\frac{\hbar}{m{\bar{\omega}}}}$, we can write this differential equation as
\begin{equation}
\left[-\frac{d^2}{dx^2}-\frac{1}{x}\frac{d}{dx}+\frac{l^2}{x^2}-2(l+1)+x^2-\frac{E^2}{\hbar\bar{\omega}mc^2}+\frac{mc^2}{\hbar\bar{\omega}}\right]G(x)=0.\label{dgx}
\end{equation}
A similar equations holds for the wave function $F(r)$. In reference \cite{Gur}, the $su(1,1)$ Lie algebra generators were introduced for the
isotropic harmonic oscillator in arbitrary dimensions. For the two-dimensional space, the $su(1,1)$ Lie algebra generators are
\begin{equation}
K_{\pm}=\frac{1}{2}\left(\pm x\frac{d}{dx}-x^2+2K_0\pm 1\right),\quad\quad K_0=\frac{1}{4}\left(-\frac{d^2}{dx^2}-\frac{1}{x}\frac{d}{dx}+\frac{l^2}{x^2}+x^2\right).
\end{equation}
The eigenfunctions basis for an irreducible unitary representations of the $su(1,1)$ Lie algebra (Sturmian basis)
in $2$-dimensional space are \cite{Gur,Moshlib}
\begin{equation}
R_{n}^l(x)=\left[\frac{2\Gamma(n+1)}{\Gamma(n+l+1)}\right]^{1/2}x^{l}e^{-x^2/2}L_{n}^{l}(x^2)\label{sturm},
\end{equation}
where $L_n^l(r^2)$ are the associated Laguerre polynomials.

Therefore, the differential equation for the upper component $G(x)$ can be written in terms of $K_0$ as
\begin{equation}
\left[4K_0-2(l+1)-\frac{E^2}{\hbar\bar{\omega}mc^2}+\frac{mc^2}{\hbar\bar{\omega}}\right]G(x)=0.\label{hk}
\end{equation}
From the action of the Casimir operator $K^2$ on the radial function $G(x)$ and the theory of unitary irreducible representation it follows that
\begin{equation}
K^2G(x)=\left(\frac{l}{2}+\frac{1}{2}\right)\left(\frac{l}{2}+\frac{1}{2}-1\right)G(x)=k(k-1)G(x).
\end{equation}
Thus, the group number $k$ (Bargmann index) has two possibilities,
\begin{equation}
k=\frac{1}{2}(l+1), \quad\quad k=-\frac{1}{2}(l-1).
\end{equation}
Since $l=0, \pm1, \pm2,...$, both choices produces discrete series $(k>0)$. Without lose of generality, in what follows
we will consider only the case for which $k=\frac{1}{2}(l+1)$. By substituting this result into equation (\ref{hk}) and by using equation (\ref{k+n}),
we obtain
\begin{equation}
K_0G(x)=\left[k+\frac{E^2}{4\hbar\bar{\omega}mc^2}-\frac{mc^2}{4\hbar\bar{\omega}}\right]G(x)=(k+n)G(x).
\end{equation}
Thus, we can identify the other group number $n$ as
\begin{equation}
n=\frac{E^2}{4\hbar\bar{\omega}mc^2}-\frac{mc^2}{4\hbar\bar{\omega}},
\end{equation}
from which we obtain the energy spectrum $E$ for the Dirac-Moshinsky oscillator coupled to an external magnetic field
\begin{equation}
E=\pm mc^2\sqrt{1+\frac{4\hbar\bar{\omega}}{mc^2}n}, \quad\quad n=0,1,2,...
\end{equation}
This result is in agreement with those previously reported in references \cite{Mandal,Boumali}. Notice that our result mainly rests on the
$SU(1,1)$ theory of unitary irreducible representations. Moreover, since the differential equation for $G(r)$ was written by means of the $su(1,1)$ Lie algebra generators, this radial component can be expressed in terms of the group basis (\ref{sturm}) as
\begin{equation}
G(r)=A_n(ar)^l e^{-a^2r^2/2}L_n^l(a^2r^2),\label{gr}
\end{equation}
where $a=b^{-1}=\sqrt{\frac{m\bar{\omega}}{\hbar}}$. It can be shown that the equation for the other radial component $F(r)$ can be obtained from equation (\ref{dgx}) by making the changes $l\rightarrow l+1$, $n\rightarrow n-1$ and $G(r)\rightarrow F(r)$. Thus, the lower Dirac radial component $F(r)$ is
\begin{equation}
F(r)=B_n(ar)^{l+1} e^{-a^2r^2/2}L_{n-1}^{l+1}(a^2r^2).\label{fr}
\end{equation}
In these last equations $A_n$ and $B_n$ are the normalization constants to be determined by following the procedure of reference \cite{UVAROV}. The relationship between these coefficients is obtained by substituting the functions (\ref{gr}) and (\ref{fr}) into the coupled equation (\ref{cou1}), and taking the limit
$r\rightarrow 0$. This leads to
\begin{equation}
B_n\left[2a(l+1)L_{n-1}^{l+1}(0)\right]=\frac{(E-mc^2)}{c\hbar}A_nL_n^l(0).
\end{equation}
The well known property of the Laguerre polynomials at the limit $r\rightarrow 0$
\begin{equation}
L_n^{l}(0)=\frac{\Gamma(n+l+1)}{n!\Gamma(l+1)},
\end{equation}
allows to simplify the relationship between $A_n$ and $B_n$ to
\begin{equation}
B_n=\frac{E-mc^2}{2nac\hbar}A_n.
\end{equation}
The relativistic normalization for a two-dimensional space is given by \cite{Tutik}
\begin{equation}
\int_0^{\infty}r\left(F^*F+G^*G\right)dr=\int_0^{\infty}r\left(F^2+G^2\right)dr=1.\label{norm}
\end{equation}
The integrals are calculated by using the orthogonality of the Laguerre polynomials $L_n^l(x)$ \cite{Ju}
\begin{equation}
\int_0^{\infty}x^{l}e^{-x}L_n^l(x)L_m^l(x)dx=\frac{\Gamma(n+l+1)}{n!}\delta_{nm}.
\end{equation}
With these results we find that the normalization constant $A_n$ is
\begin{equation}
A_n^2=\frac{2a^2n!}{\Gamma(n+l+1)\left[1+n\left(\frac{E-mc^2}{2anc\hbar}\right)^2\right]}.
\end{equation}
Therefore, the radial functions $G(r)$ and $F(r)$ for the Dirac-Moshinsky oscillator coupled to an external
magnetic field are given explicitly by
\begin{equation}
\begin{pmatrix}
G(r)\\
F(r)
\end{pmatrix}=A_n(ar)^le^{-\frac{a^2r^2}{2}}
\begin{pmatrix}
L_{n}^{l}(a^2r^2)\\
\left(\frac{E-mc^2}{2nc\hbar}\right)rL_{n-1}^{l+1}(a^2r^2)
\end{pmatrix}.\label{radiales2}
\end{equation}
Hence, we solved the problem in an new and elegant way by using pure algebraic methods. We emphasize two important facts. First, the redefinition
of the angular frequency $(\omega\rightarrow\bar{\omega})$ lets the algebraic structure of the Dirac-Moshinsky oscillator unchanged. Second, each of the decoupled radial equations admits an $su(1,1)$ algebraic treatment. In the next section we shall obtain advantage of this point by constructing the relativistic coherent state for this problem.

\section{$SU(1,1)$ radial coherent states}

A summary of the theory of coherent states is presented in the Appendix. Since the radial functions of this problem $G(r)$ and $F(r)$ are obtained from the group basis (Sturmian basis), we can construct their coherent states $G(r,\xi)$ and $F(r,\xi)$. By substituting equations (\ref{gr}) and (\ref{fr}) into equation (\ref{PCN}) we obtain
\begin{equation}
G(r,\xi)=\left[\frac{2(1-|\xi|^2)^{l+1}}{\Gamma(l+1)}\right]^{\frac{1}{2}}(ar)^le^{-\frac{a^2r^2}{2}}\sum_{n=0}^{\infty}\xi^nL_n^l(a^2r^2),
\end{equation}
\begin{equation}
F(r,\xi)=\left[\frac{2(1-|\xi|^2)^{l+2}}{\Gamma(l+2)}\right]^{\frac{1}{2}}(ar)^{l+1}e^{-\frac{a^2r^2}{2}}\sum_{n=1}^{\infty}\xi^{n-1}L_{n-1}^{l+1}(a^2r^2).
\end{equation}
If we use the generating function for the Laguerre polynomials
\begin{equation}
\sum_{n=0}^\infty L_n^\nu(x)y^n=\frac{e^{xy/(1-y)}}{(1-y)^{\nu+1}}, \quad\quad |y|<1,
\end{equation}
the sums can be calculated in a closed form and the coherent states $G(r,\xi)$ and $F(r,\xi)$ result to be
\begin{equation}
G(r,\xi)=A_n'\frac{(ar)^l}{(1-\xi)^{l+1}}e^{\frac{a^2r^2-3a^2r^2\xi}{2(1-\xi)}},\quad\quad
F(r,\xi)=B_n'\frac{(ar)^{l+1}}{(1-\xi)^{l+2}}e^{\frac{a^2r^2-3a^2r^2\xi}{2(1-\xi)}},\label{fgc}
\end{equation}
where $A_n'$ and $B_n'$ are two new normalization constants. The relationship between the normalization coefficients $A_n'$ and $B_n'$ can be obtained by noting that the coherent states $G(r,\xi)$ and $F(r,\xi)$ must satisfy the coupled equation (\ref{cou1}). Consequently, the substitution of equations (\ref{fgc}) into equation (\ref{cou1}), and considering the limit $r\rightarrow 0$, leads to
\begin{equation}
B_n'=\frac{E-mc^2}{c\hbar a(l+1)}A_n'.
\end{equation}
The coefficient $A_n'$ is obtained from the relativistic normalization, equation (\ref{norm}). The two integrals
are reduced to an appropriate Gamma function. Thus $A_n'$ results to be
\begin{equation}
A_n'^2=\frac{(1-\xi)^{2(l+1)}}{1+\left(\frac{E-mc^2}{ac\hbar(l+1)(1-\xi)}\right)^2}.
\end{equation}
With these previous results, we are in position to give the explicit form of the $SU(1,1)$ relativistic coherent state for the
Dirac-Moshinsky oscillator coupled to an external magnetic field
\begin{equation}
\begin{pmatrix}
G(r,\xi)\\
F(r,\xi)
\end{pmatrix}=A_n'\frac{(ar)^l}{(1-\xi)^{l+1}}e^{\frac{a^2r^2-3a^2r^2\xi}{2(1-\xi)}}
\begin{pmatrix}
1\\
\frac{E-mc^2}{c\hbar a(l+1)(1-\xi)}
\end{pmatrix}.\label{radialescoherentes}
\end{equation}

The coherent states of Dirac-Moshinsky oscillator were first introduced with the study of the behavior of wave packets in $(1+1)$ dimensions \cite{Nogami}. These relativist coherent states are related to the Heisenberg-Weyl algebra. The relativistic coherent states were also calculated for the
Dirac-Kepler-Coulomb problem in $(3+1)$-dimensions \cite{Draga} and its generalization to $(D+1)$-dimensions \cite{Didier}.

\section{Concluding remarks}

We have studied the $(2+1)$-dimensional Dirac-Moshinsky oscillator coupled to an external magnetic field
by using pure algebraic methods. The eigenfunctions and energy spectrum were obtained by properly introducing
the $su(1,1)$ Lie algebra theory and its irreducible unitary representation to the Dirac radial equations.
Moreover, the $SU(1,1)$ group theory and its Sturmian basis allowed us to find the radial
coherent states for each spinor component. We used relativistic normalization to obtain the final form
of the $SU(1,1)$ relativistic radial coherent states for this problem. We emphasize that the method used in this work
to obtain the eigenfunctions, the eigenspectrum and the radial coherent states for this problem is purely algebraic.

\section*{Acknowledgments}
This work was partially supported by SNI-M\'exico, COFAA-IPN,
EDI-IPN, EDD-IPN, SIP-IPN project number $20140598$.

\section{Appendix}

Three operators $K_{\pm}, K_0$ close the $su(1,1)$ Lie algebra if they satisfy the commutation relations \cite{Vourdas}
\begin{eqnarray}
[K_{0},K_{\pm}]=\pm K_{\pm},\quad\quad [K_{-},K_{+}]=2K_{0}.\label{com}
\end{eqnarray}
The action of these operators on the Fock space states
$\{|k,n\rangle, n=0,1,2,...\}$ is
\begin{equation}
K_{+}|k,n\rangle=\sqrt{(n+1)(2k+n)}|k,n+1\rangle,\label{k+n}
\end{equation}
\begin{equation}
K_{-}|k,n\rangle=\sqrt{n(2k+n-1)}|k,n-1\rangle,\label{k-n}
\end{equation}
\begin{equation}
K_{0}|k,n\rangle=(k+n)|k,n\rangle,\label{k0n}
\end{equation}
where $|k,0\rangle$ is the lowest normalized state. The Casimir
operator $K^{2}$ for any irreducible representation of this group is given by
\begin{equation}
K^2=K^2_0-\frac{1}{2}(K_+K_-+K_-K_+)
\end{equation}
and satisfies the relationship $K^{2}=k(k-1)$. Thus, a representation of
$su(1,1)$ algebra is determined by the number $k$, called the Bargmann index. The discrete series
are those for which $k>0$.

The $SU(1,1)$ Perelomov coherent states are defined as the action of the displacement operator $D(\xi)$
onto the lowest normalized state $|k,0\rangle$ as \cite{Perellibro}
\begin{equation}
|\zeta\rangle=D(\xi)|k,0\rangle=(1-|\zeta|^2)^k\sum_{n=0}^\infty\sqrt{\frac{\Gamma(n+2k)}{n!\Gamma(2k)}}\zeta^n|k,n\rangle,\label{PCN}
\end{equation}
The displacement operator $D(\xi)$ is defined in terms of the creation and annihilation operators $K_+, K_-$ as
\begin{equation}
D(\xi)=\exp(\xi K_{+}-\xi^{*}K_{-}),\label{do}
\end{equation}
where $\xi=-\frac{1}{2}\tau e^{-i\varphi}$, $-\infty<\tau<\infty$ and $0\leq\varphi\leq2\pi$.
The so-called normal form of the squeezing operator is given by
\begin{equation}
D(\xi)=\exp(\zeta K_{+})\exp(\eta K_{0})\exp(-\zeta^*K_{-})\label{normal},
\end{equation}
where  $\zeta=-\tanh(\frac{1}{2}\tau)e^{-i\varphi}$ and $\eta=-2\ln \cosh|\xi|=\ln(1-|\zeta|^2)$ \cite{Gerry}.

\end{document}